\documentclass[%
aps,
prl,
superscriptaddress,
amsmath,amssymb,
 reprint,%
]{revtex4-1}
\usepackage{color}
\usepackage{graphicx}
\usepackage{dcolumn}
\usepackage{bm}
\usepackage{booktabs}
\usepackage[utf8]{inputenc}
\usepackage[T1]{fontenc}
\usepackage{mathptmx}
\begin{document}
	
	
	\title{Generating axial magnetic fields via two plasmon decay driven by a twisted laser}

\author{Yu Ji}
\affiliation{ 
	Department of Modern Mechanics, University of Science and Technology of China, Hefei 230026, China
}%
\author{Chang-Wang Lian}
\affiliation{Department of Plasma Physics and Fusion Engineering, University of Science and Technology of China, Hefei, Anhui 230026, China}
\affiliation{
	Laser Fusion Research Center, China Academy of Engineering Physics, Sichuan, Mianyang, 621900, China}
\author{Yin Shi}
\affiliation{Department of Plasma Physics and Fusion Engineering, University of Science and Technology of China, Hefei, Anhui 230026, China}
\author{Rui Yan}%
\email{ruiyan@ustc.edu.cn}
\affiliation{ 
	Department of Modern Mechanics, University of Science and Technology of China, Hefei 230026, China
}%
\affiliation{Collaborative Innovation Center of IFSA (CICIFSA), Shanghai Jiao Tong University, Shanghai 200240, China}
\author{Shihui Cao}
\affiliation{Department of Mechanical Engineering, University of Rochester, Rochester, New York 14627, USA}%
\author{Chuang Ren}
\affiliation{Department of Mechanical Engineering, University of Rochester, Rochester, New York 14627, USA}%
\affiliation{Department of Physics and Astronomy, University of Rochester, Rochester, New York 14627, USA}%
\author{Jian Zheng}
\affiliation{Department of Plasma Physics and Fusion Engineering, University of Science and Technology of China, Hefei, Anhui 230026, China}
\affiliation{Collaborative Innovation Center of IFSA (CICIFSA), Shanghai Jiao Tong University, Shanghai 200240, China}


\date{\today}
\begin{abstract}
We propose a new way of axial magnetic fields generation in a non-relativistic laser intensity regime by using a twisted light carrying orbital angular momentum (OAM) to stimulate two-plasmon decay (TPD) in a plasma. The growth of TPD driven by an OAM light in a Laguerre-Gauss (LG) mode is investigated through three dimensional fluid simulations and theory. A theory based on the assumption that the electron plasma waves (EPWs) are locally driven by a number of local plane-wave lasers predicts the maximum growth rate proportional to the peak amplitude of the pump laser field and is verified by the simulations. The OAM conservation during its transportation from the laser to the TPD daughter EWPs is shown by both the theory and the simulations. The theory predicts generation of $\sim$ 40T axial magnetic fields through the OAM absorption via TPD, which has perspective applications in the field of high energy density physics.



\end{abstract}

\maketitle

Absorption of angular momentum from an intense twisted laser that carries orbital angular momentum (OAM)\cite{Allen2000} can lead to the generation of strong axial magnetic fields in a plasma. This so-called inverse Faraday effect for OAM lasers\cite{Ali2010} has recently attracted intensive research interest in relativistic-intensity ($I \gtrsim 10^{18}\rm{W/cm^2}$ ) laser plasma interactions\cite{Longman2021,Nuter2020,Shi2018}. At relatively lower intensities ($I \sim 10^{14}-10^{16}\rm{W/cm^2}$ )  relevant to the leading facilities for inertial confinement fusion (ICF), an OAM laser is capable of passing its OAM selectively to electrons or ions in a plasma, in this regime via the laser plasma instability (LPI) processes\cite{Mendonca2009, Vieira2016,Nuter2022,Gao2015,Feng2022}, while few work has been reported on the axial magnetic field generation. When getting absorbed and decaying into a pair of daughter waves, an OAM laser is expected to transport OAM to its daughter waves via three-wave coupling due to the angular momentum conservation. Although both an electron plasma wave (EPW)\cite{Mendonca2009a,Blackman2019,Blackman2019a,Blackman2022} and an ion-acoustic wave (IAW)\cite{Ayub2011} can appear as a Laguerre-Gauss (LG) mode that carries OAM in a plasma, the OAM absorption from a laser into such a wave through LPI is not that efficient.
The OAM transportation in paraxial LPI scenarios for Stimulated Raman/Brillouin Scattering (SRS/SBS), whose two daughter waves include a scattered light (electromagnetic) wave and an EPW/IAW has been analytically modeled and numerically simulated\cite{Mendonca2009, Vieira2016,Feng2022,Shi2018}. In these studies, to ensure angular momentum being passed from an LG pump laser to an EPW or IAW, the daughter light wave need to be seeded in an LG mode that already carries OAM and is launched collimated with the axle of the pump laser. 
The other EPW/IAW daughter wave in a paraxial LG form is then beaten  by the collimated pump and scattered light satisfying the matching conditions on frequencies, wave vectors, and the azimuthal indices representing OAM \cite{Mendonca2009, Vieira2016,Feng2022,Shi2018}. However, SRS driven from thermal noises without such an LG-form seed demonstrates that virtually no angular momentum is transported to the daughter EPW while only the scattered light wave carries OAM\cite{Nuter2022}. 

These pioneering studies prompt Two Plasmon Decay (TPD) as an efficient candidate favoring the OAM absorption into the electrons of a plasma. TPD is a fundamental LPI process occurring when an electromagnetic wave decaying into a pair of EPWs \cite{Liu1976,Kruer2019physics} in the region where the electron number density is below 1/4 of the critical density ($n_{cr}$) above which a laser is not able to propagate through. TPD is identified as a critical concern in ICF for its low threshold\cite{Liu1976,Simon1983,Lian2022} and fuel-preheating risk caused by the energetic electrons it generates. Since both daughter waves of TPD are EPWs, the OAM carried by the pump laser is able to go nowhere else but into the daughter TPD EPWs due to the angular momentum conservation. TPD is intrinsically excited in a non-collimated way as the most unstable modes involve pairs of EPWs that propagate at fairly large angles with respect to the propagation direction of the pump laser\cite{Kruer2019physics}. This non-collimated geometry of the dominant modes is another key feature that would make TPD grow in a very distinguished way from the collimated forward/backward SRS and SBS in the previous OAM-SRS/SBS studies\cite{Mendonca2009, Vieira2016,Feng2022,Nuter2022}.   

In this Letter, we for the first time present the fluid simulations and theory on TPD growth driven by an OAM laser with moderate azimuthal indices in a homogeneous plasma, and propose a new method generating axial magnetic fields through TPD. We focus on a most common scenario where an LG pump laser shines in a plasma and stimulates the TPD EPWs naturally from random noises. The laser intensities $(I \thicksim 10^{15} \rm{W/cm^{2}})$ are relevant to ICF experimental conditions.  
The TPD EPWs are found to propagate at large angles with the axle of the incident laser and can be recognized approximately as a collection of enormous local planar plasma waves. A theory is then developed by decomposing the LG laser into local tilted plane waves which locally drive TPD EPWs via three-wave coupling. The TPD growth rate predicted by the theory matches our simulation results very well for different azimuthal indices of the incident laser. This theory also provides a prediction of the angular momentum collectively carried by the TPD EPWs, which excellently agrees with the simulation results and verifies the angular momentum conservation in such a non-collimated LPI system. The TPW EPWs collectively form a spiral current tube that generates axial magnetic fields whose magnitude is theoretically estimated.
Thanks to the encouraging progresses of the laser technology\cite{Wang2020,Gao2015}, OAM lasers at relevant intensities and an experimental validation could be expected in the near future.


We have performed a series of three-dimensional(3D) fluid simulations using our newly developed code \textit{FLAME-MD} \cite{Zhou2022}. In \textit{FLAME-MD} the set of fluid-like equations presented in Ref. \cite{Hao2017} is solved in 3D space without taking envelopes in either space or time. The pump LG laser propagating along the $z$ direction is prescribed in the vector potential in an LG form of

\begin{equation}
	\begin{aligned}
		{\textbf A}&(x,y,z,t) = {\frac{1}{2}} {\textbf A_{0}}[\frac{r\sqrt{2}}{w_{b}(z)}]^{|l|} L^{|l|}_{p}(-{\frac{2r^{2}}{w_{b}^{2}(z)}}) \exp(-{\frac{r^{2}}{w_{b}^{2}(z)}})\\ 
		 \times & \exp[i\omega_0t-ik_0(z-z_0)+{\frac{ik_{0}(z-z_0)}{1+(z-z_0)^{2}/z_R^{2}}}{\frac{r^{2}}{z_R^{2}}}\\  & -i(2p+|l|+1)\arctan({\frac{z-z_0}{z_R}}) +i \theta_{0} +i l \phi]+ c.c.,
	\end{aligned}
	\label{eq:lg-A-profile}
\end{equation}
where ${\textbf A_{0} }$ is the vector potential at the focus, ${(x_0,y_0,z_0)}$ is the coordinate of the center of focus, $r\equiv \sqrt{(x-x_0)^2+(y-y_0)^2}$ is the radial distance to the laser axle, $w_{b}^{2}(z) \equiv w^{2}_{b0}[1+(z-z_0)^{2}/z_R^{2}]$ is the beam waist, $w_{b0}$ is the waist on the focal plane,  $z_R$ is the Rayleigh length, 
$\omega_0$ and $k_0$ are its central frequency and wavenumber in plasma. ${L^{|l|}_{p}}$ is a generalized Laguerre Polynomial of order ${(p,l)}$ with ${\textit l}$ known as the azimuthal index that gives rise to OAM and $p$ the radial index that controls the number of zeros along the radial direction. In this Letter  we only consider the most commonly occurring $p=0$  modes. ${\theta_{0}}$ is an initial phase and ${\phi}$ is an azimuthal angle. The complex conjugate (c.c.) is added to make $\textbf{A}$ real.

A typical simulation box is  $22 {\rm \mu m}$(x) $\times$ $22 {\rm \mu m}$ (y) $\times 4.5 \mu$m (z) with the grid of $2000 \times 2000 \times 400$. The LG laser is focused at the center of box with the waist width ${\textit w_{b0}} = 2.8{\rm \mu m}$. The LG laser whose central wavelength in vacuum is $\lambda 
= 0.351\mu$m  is linearly polarized along the $y$ direction. A longer Rayleigh length ($z_R= 245\mu$m) than physical for this spot size is set to mitigate the laser focusing effects near its waist while accommodating the computational costs. The module in\textit{ FLAME-MD} solving for the scattered lights is intentionally turned off to ensure that TPD is the only possible LPI process in this scenario while SRS and SBS involving scattered lights are naturally inhibited. 
A uniform electron density $n_0 = 0.23 n_{cr}$ is set with the electron temperature $T_e = 1.5\rm{keV}$. The ions are fixed and Landau damping is turned off, to make sure that TPD is growing in its linear (exponentially growing) phase without nonlinear saturation mechanisms involved.

A series of simulations with moderate $l$ ranging from 0 to 6 are performed. Since $p=0$, the case with $l=0$ retreats to a regular Gaussian laser carrying no OAM. For a fair comparison on TPD growth that was found sensitive to the local maximum electric field of a laser\cite{Lian2022}, the maximum vector potential's magnitude ($A_{max}$) normalized as $a_{max} = e A_{max} /(m_e c^2) = 0.024$ is set identical for different $l$ cases. Here $e$ is the electron charge, $m_e$ is the electron mass and $c$ is the light speed in vacuum. $a_{max} = 0.024 $ is associated with the  intensity $I_{max} = 6\times 10^{15} /rm{W/cm^{2}}$ for an ICF-relevant Gaussian laser via $a_{max} = 0.0085\times \sqrt{I_{14}^{max} \lambda_{\mu m}^2}$, where $I_{14}^{max}$ is $I_{max}$ in $10^{14}\rm{W/cm^2}$, $\lambda_{\mu m}$ is $\lambda$ in $\rm{\mu m}$, respectively.
\begin{figure}
	\includegraphics[width=3.375in]{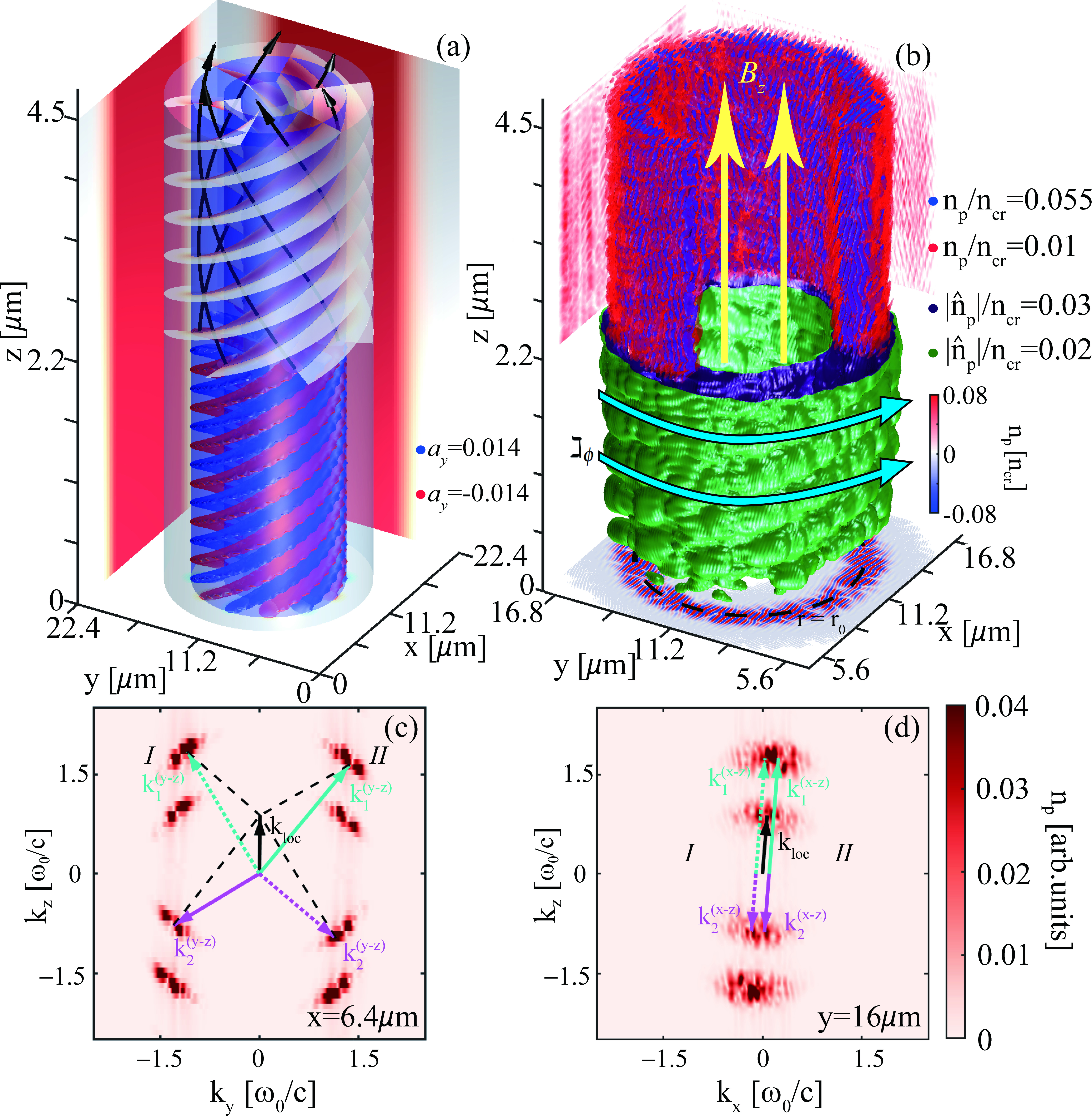}
	\centering
	\caption{Set-up of the LG laser with $l=6$ (a) and the electron density modulations ($n_p$) at $t=0.75\rm{ps}$(b-d). (a)(upper half) Phase iso-surfaces (white helical fronts) and the schematic of local Poynting vectors' directions (black arrows). (lower half) Iso-surfaces of the normalized vector potential's $y$ component ($a_y=e A_y/m_ec^2$) at $|a_y|^2=2.56\times 10^{-6}$ (outer white cylinder) and $2.34\times 10^{-4}$ (inner blue cylinders). (b)(upper half) $n_p$ iso-surfaces with a quarter segment cut away. (lower half) Enveloped electron density modulation ($\hat{n}_p$) iso-surfaces. The x-y projection shows the slice of $n_p$ on the focal plane of the laser, overlaid by the dashed circle $r=r_0$ where  $a_{max}$ is reached. 
		The time-averaged azimuthal current and the axial magnetic fields induced are marked by the blue and yellow arrows, respectively. 
		(c)-(d) Spectra of $n_p$ on a $y-z$ slice (c) at $x=6.4\rm{\mu m}$ [passing Point A of Fig.\ref{fig:spec}(a)] and on a $x-z$ slice (d) at $y=16\rm{\mu m}$ [passing Point C of Fig.\ref{fig:spec}(a)]. The arrows illustrate the TPD wavevector matching geometry projected on the slices for two pairs of dominant EPWs[I (dotted) and II(solid)]. The black solid arrows are the local laser wave vectors  $\textbf{k}_{loc}$.
	}
	\label{fig:scope}
\end{figure}

The contour of $a$ of the pump laser in the $ {\textit l} = 6$ case is illustrated in the lower half of Fig.\ref{fig:scope}(a) while the vortex property of the local Poynting vectors perpendicular to the helical wave fronts are marked as arrows in the upper half of Fig.\ref{fig:scope}(a). The spiral directions of the Poynting vectors exhibit the OAM along $z$. 
A global scope of the EPWs stimulated in the same case is illustrated in Fig.\ref{fig:scope}(b). The upper half shows electron density modulations due to EPWs ($n_p$) and the lower half shows an envelope of $n_p$. It is shown that the EPWs are concentrated on a hollow cylinder near $r=r_0\equiv w_{b0}\sqrt{|l|/2}$ where the pump laser reaches its radial peak intensity of the LG profile ($r_0 = 4.8\rm{\mu m}$ for the case  $ {\textit l} = 6$). The spectra of $n_p$ on two slices tangent to the cylinder of $r=r_0$ are shown in Fig.\ref{fig:scope}(c) and (d). The dominant EPWs are found to propagate at large angles with respect to the laser propagation direction on the $y-z$ slice where the laser is polarized [see Fig.\ref{fig:scope}(c)] within this plane, while the dominant EPWs are concentrated at small $k_x$'s on the $x-z$ slice [see Fig.\ref{fig:scope}(d)].
These features which are largely at odds with a paraxial LG mode, are similar to the features in the simulations with a plane-wave laser\cite{Yanthesis2012,Wen2016}.

\begin{figure}
	\includegraphics[width=3.375in]{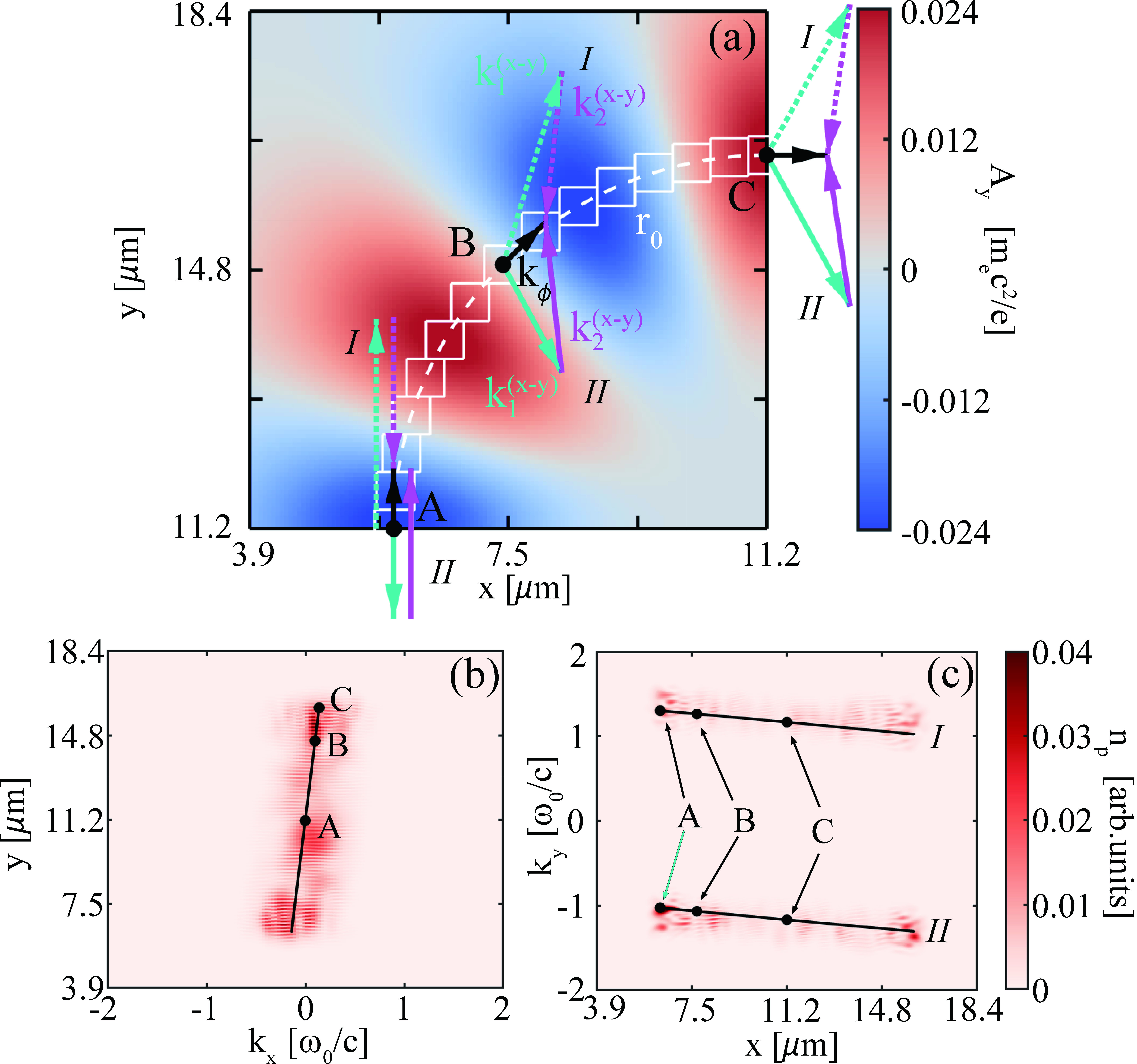}
	\centering
	\caption{(a) The LG laser vector potential's y components ($A_y$) on its focal plane in the simulation with $l=6$ at $t=0.75$ps. The white squares are the schematic of the zones $S_n$ on the cylinder $r_n=r_0$ (dashed line). The arrows are the $k_x-k_y$ plane projections of ${\textbf k_{loc}}$(black), ${\textbf k_1}$(purple) and ${\textbf k_2}$(cyan) at three locations (A, B, and C), respectively. The dashed and solid arrows at A, B, and C show the wavevector matching geometry of the two pairs of dominant EPWs ($I,II$), respectively. 
		(b)-(c) Spectra of $n_p$ in $k_x-y$ (b) and  $k_y-x$ (c) spaces overlaid by the theoretical prediction of the dominant EPWs' $\textbf{k}_1$ using Eq.\ref{eq:rotation} (solid line) with the values at A, B, and C marked in black dots. Only the EPWs whose wavevector is $\textbf{k}_1$ are shown in the spectra.
	}	
	\label{fig:spec}
\end{figure}

To model the growth of TPD driven by an LG laser with arbitrary OAM states,  we have developed a new theory based on the EPWs' features shown in our simulations. The EPWs originated from thermal noises can be approximately recognized as a number of local planar waves stimulated by local plane-wave pumps via TPD. Correspondingly, we can divide the space into numerous small zones ($S_n$, n= 1,2,...) in each of which the piece of the LG laser is approximated by a plane wave. $S_n$ looks like a fine rod that is parallel to the laser axle, ie. the $z$-axis, and has a small cross section centered at ($x_n$, $y_n$) on the $x$-$y$ plane. The zones on $r=r_0$ are illustrated by the squares in Fig.\ref{fig:spec}(a). The laser field for $p=0$ (the Laguerre polynomial ${L^{|l|}_{p}}$ is unity when ${p =0}$) on the large Rayleigh length limit ($z \ll z_R$) can be piecewisely described by Taylor expanding Eq. (\ref{eq:lg-A-profile}) on ($x_n, y_n$) in each individual zone, ie.,
\begin{equation}
	\begin{aligned}
		&{\textbf A(x,y,z,t)} = \sum\limits^n {\frac{1}{2}} {\textbf A_n}  T_n(x',y')(1+P_{xn} x' + P_{yn} y')\\
		&\exp [i k_{yn} y' + i k_{xn} x'-ik_0(z-z_0) ]+O({x'}^2,{y'}^2)+c.c.,
		\label{eq:LG-series}
	\end{aligned}
\end{equation}
where ${\textbf A_n}\equiv {\textbf A}(x_n, y_n,z_0,t)$, $x'\equiv x-x_n,y' \equiv y-y_n$, $T_n$ is a flat-top window function such that $T_n(x',y')=1$ when $(x,y,z) \in S_n$ and 0 otherwise,  $P_x = x_n(lw_{b0}^2-2r_n^2)/(w_{b0}^2r_n^2)$, $P_y = y_n(lw_{b0}^2-2r_n^2)/(w_{b0}^2r_n^2)$,  $r_n = \sqrt{(x_n-x_0)^2+(y_n-y_0)^2}$ is the radius of each zone's center with respect to the laser axle, $k_{xn} =\sin\phi_{n}l/r_{n}$, $k_{yn} =-\cos\phi_{n}l/r_{n}$, the azimuthal angle $\phi_n = \arctan[(y_n-y_0)/(x_n-x_0)]$ is $\phi$ in Eq.(\ref{eq:lg-A-profile}) evaluated at ($x_n, y_n$). The leading terms of Eq.(\ref{eq:LG-series}) sketch inside each zone a local plane wave propagating at $\textbf{k}_{loc} = (k_{xn},k_{yn},k_0)^T \equiv \textbf{k}_{\phi}+ \textbf{k}_{0}$, which is illustrated by the wavevector-matching-condition triangles in Fig.\ref{fig:scope}(c) and (d).


As EPWs are assumed locally driven via TPD in each zone in this model, the fastest growing modes which we are most interested in are expected to be determined by the zones [see Fig.\ref{fig:spec}(a)] illuminated at the highest laser intensity, ie. $r_n=r_0$ and ${\textbf A_{n}} = {\textbf A_{max}} = {\textbf A_0}\exp[-|l|/2]|l|^{(|l|/2)}$. It is straightforward to find $P_{xn}=P_{yn}=0$ if $r_n=r_0$, indicating a uniform laser intensity in a zone to the first-order accuracy. The problem finding the growth of the leading modes then retreats to a problem finding a number of local plane-wave TPD scenarios inside each zone on the cylinder $r_n=r_0$. Inside such a zone, the differential equations governing TPD growth (Eqs. 1. 2. and 3 in Ref. \cite{Yan2010}) can be cast to a three-wave interaction model following the procedures in Ref. \cite{Yan2010} and \cite{Yan2009} as $[\partial_t +v_{z1,z2} \partial_z]a_{1,2}=\gamma_0 a_{2,1}^*$,
where the subscripts $1,2$ denote a pair of daughter EPWs that satisfy the matching conditions $\textbf{k}_{loc} = \textbf{k}_1 + \textbf{k}_2$, and $\omega_0=\omega_1+\omega_2$, $\omega_{1,2},{\textbf k_{1,2}}$ are the EPWs' frequencies and wave vectors, respectively. $v_{z1,z2}$ are the $z$ components of the EPWs' group velocities, $a_{1,2}\equiv \alpha_{1,2} {\Phi}_{1,2}  $ are the enveloped amplitudes of the two daughter EPWs, $\alpha_{1,2} \equiv \sqrt{|k_{y1,y2}v_0||\omega_{1,2}/\omega_{2,1}-k_{1,2}^2/k_{2,1}^2|/4}$, $k_{y1,y2}$ are the $y$ components of ${\textbf k_{1,2}}$, $\Phi_{1,2}$ are the enveloped electrostatic potentials \cite{Yan2010}, and ${\textbf v_{0}=e{\textbf A_{max}}/m_e c}$ is usually known as the electron oscillation velocity in a laser field. Without losing generality, the higher-frequency EPW which also has a larger wave number of the paired daughter EPWs are denoted by subscript 1. The TPD growth rate is given by the coupling coefficient $\gamma_0$:
\begin{equation}
	\begin{aligned}
		\gamma_0&=\sqrt{\frac{1}{16}k_\bot^2 v_0^2(\frac{\omega_2}{\omega_1}-\frac{k_2^2}{k_1^2}) (\frac{\omega_1}{\omega_2}-\frac{k_1^2}{k_2^2})},
	\end{aligned}
	\label{eq:gamma0}
\end{equation}
where $k_\bot \equiv |k_{y1}-k_{y2}|/2$. The factor $k_\bot^2$ in Eq.(\ref{eq:gamma0}) has been  approximated from $|k_{y1}k_{y2}|$ by neglecting the second-order term of $k_{yn}$ as $|k_{y1}k_{y2}|=k_\bot^2-k_{yn}^2/4$ and $|k_{yn}|\ll k_0$. The dominant EPW pairs which maximize $\gamma_0$ can be found in each zone using Eq.(\ref{eq:gamma0}) and the matching conditions. The same maxima of $\gamma_0$ in all zones on $r=r_0$ are expected to be reached at the same values of $k_1$, $k_2$, and $k_\bot^2$, while different $\textbf{k}_{\phi}$ [see Fig.\ref{fig:spec}(a)] causes different directions of $\textbf{k}_{1}$ in each zone. The dominant $\textbf{k}_{1}$ is located within the plane determined by $\textbf{k}_{loc}$ and the polarization direction (ie. the $y$ axis). 

The dominant $\textbf{k}_{1}$ in each zone can be considered as a rotation of $\textbf{k}_{norm}$, which is the dominant $\textbf{k}_{1}$ driven by a plane-wave laser polarized along $y$ and propagating along $z$ at the wavevector $\sqrt{k_0^2+k_\phi^2}$. $\textbf{k}_{norm}$ can be numerically found as $\textbf{k}_{norm} = (0, \pm 1.23,1.75)^T\omega_0/c$ for the particular $n_e$ and $T_e$ in our cases. $\textbf{k}_{norm}$ has two values since a plane-wave laser drives two pairs of EPWs symmetrically on both sides. The rotation can be described as ${\textbf k_{1}} = \textbf{M}_{xy} \cdot {\textbf k_{norm}} $, where $\textbf{M}_{xy}$ is the rotation tensor written as a product of two matrices
\begin{equation}
		\textbf{M}_{xy}= \left[
		\begin{array}{ccc}
			\cos\varphi_{1} & 0 & \sin\varphi_{1} \\0 & 1 & 0 \\-\sin\varphi_{1} &0 & \cos\varphi_{1}	
		\end{array}
		\right]
		\left[
		\begin{array}{ccc}
			1 & 0            & 0             \\
			0 & \cos\varphi_{2} & \sin\varphi_{2} \\
			0 & -\sin\varphi_{2}  & \cos\varphi_{2}
		\end{array}
		\right],
	\label{eq:rotation}
\end{equation}
${\varphi_{1} = \arctan(k_{xn} /k_0)}$ is the angle between ${\textbf k_{loc}}$ and the $x$ axis, and ${\varphi_{2} = \arcsin(k_{yn}/\sqrt{k_{xn}^2+k_0^2})}$ is  the angle between ${\textbf k_{loc}}$ and its projection on the x-z plane. 

A very good agreement between the theory and the simulation on the dominant EPW's $\textbf{k}_1$ is shown in Fig.\ref{fig:spec}(b) and (c). The theory-predicted dominant $\textbf{k}_1$ by rotating $\textbf{k}_{norm}$ is plotted on top of the $n_p$ spectra from the simulation for comparison. The rotation of the two values of $\textbf{k}_{norm}$ leads to two dominant pairs of $\textbf{k}_{1}$ and $\textbf{k}_{2}$ that satisfy $\textbf{k}_{loc} = \textbf{k}_1 + \textbf{k}_2$, as illustrated by the vectors in Fig.\ref{fig:scope}(c-d) and Fig.\ref{fig:spec}(a) marked with $I$ and $II$.
Since the dominant $\textbf{k}_1$ and $\textbf{k}_2$ have distinguished modules, the EPWs whose wave vector is $\textbf{k}_2$ are filtered out and only the $\textbf{k}_1$ branches are shown in the $n_p$ spectra of Fig.\ref{fig:spec}(b) and (c). 

\begin{figure}
	\includegraphics[width=3.375in]{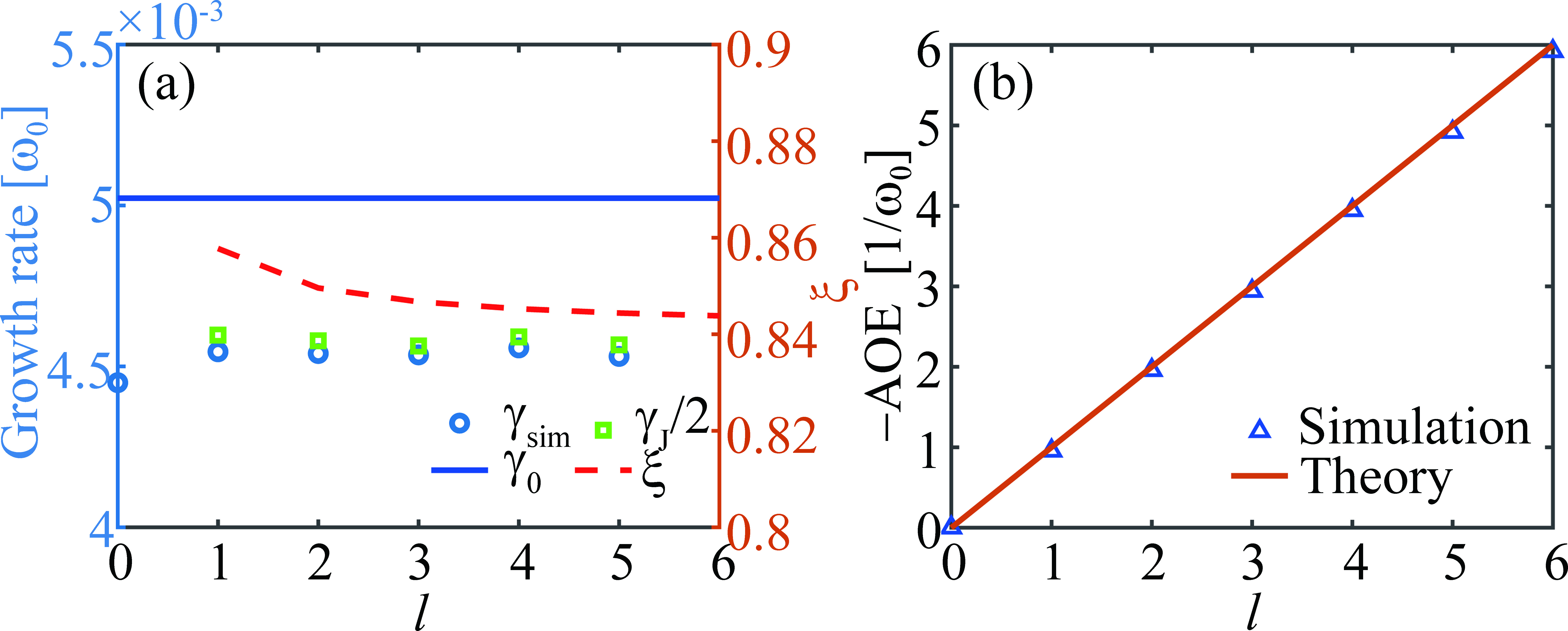}
	\centering
	\caption{(a)(left) TPD growth rate under the same $ A_{max}$ given by Eq.\ref{eq:gamma0} (blue line) and simulations (blue circles) with different $l$. Green squares are the half of growth rate for the TPD EPWs' angular momentum ($\gamma_J/2$) in simulations. (right) $\xi$ from Eq.\ref{eq:xi} for different $l$. (b) The opposite value of $AOE$ from the TPD EPWs in the theory (red line) and simulations (blue triangles) for different $l$.
	}
	\label{fig:compare}
\end{figure}

Substituting the dominant EPWs' information into Eq.(\ref{eq:gamma0}) yields the maximum of $\gamma_0 = 0.005\omega_0$ which does not depend on $l$ but only on $A_{max}$. In Fig.\ref{fig:compare}(a)  $\gamma_0$ is compared with $\gamma_{sim}$ that is measured as half of the growth rate of the volume integral of  $n_p^2$ over the entire simulation domain in a series of simulations with different $l$ but same $A_{max}$. Although $\gamma_{sim}$ is contributed by all TPD modes with different growth rates, it is determined by the dominant modes with the fastest growth after adequately long time in this linearly-growing system. Good agreement is shown in Fig.\ref{fig:compare}(a) as the relative error on the growthrate is within 10\% between the theory and the simulations. The weak dependence of $\gamma_{sim}$ on $l$ also confirms that $A_{max}$ is the key factor determining TPD growths.
The power of an LG beam varies in terms of $l$ if ${\textbf A_{max}}$ is set constant and vice versa.  
The ratio of $A_{max}$ between an LG laser and a Gaussian laser with an identical energy flux and under the same plasma conditions $\xi(l) \equiv A_{max}^{LG}/A_{max}^{Gauss}$ can be used to evaluate the TPD mitigating efficacy by using an LG laser. $\xi$ can be readily found using Eq. (\ref{eq:lg-A-profile}) as
\begin{equation}
	\xi(l) =  \sqrt{\frac{\sqrt{\pi}|l|^{|l|}}{\Gamma(|l|+\frac{1}{2})}}  \exp(-\frac{|l|}{2}),
		\label{eq:xi}
\end{equation}
when $l \neq 0 $. $\xi(0) = 1$ by definition. The values of $\xi $ at $l = 1,2,...6$ are plotted in Fig.\ref{fig:compare}(a). It is shown that the TPD growth rate is reduced by roughly 15\% if the pump light is switched from a Gaussian beam ($l=0$) to an LG beam ($l=1$) and the growth rate is slightly smaller with larger $l$ within this range of $l$ (1-6) while maintaining the same laser power. Therefore, an LG laser can be considered as a candidate of the pump in direct-drive ICF for the sake of TPD mitigation. 

The TPD EPWs collectively carry an angular momentum along the pump laser's propagation ($z$) direction. 
Again consider the fastest growing EPWs located in the zones near $r=r_0$ which will dominate the angular momentum in this exponentially-growing system. The angular momentum density ($\mathbf{j}_n$) of the EPWs in the $n$th zone can be expressed as ${\textbf j_{n}}={\textbf r_n}\times  \textbf{p}_n$, where ${\textbf p_{n}}=m_e \left \langle n_{p} {\textbf v_e} \right \rangle$ \cite{Bliokh2022} is the time-averaged (marked as $\left \langle \right \rangle$) electron momentum density at the microscopic velocity ${\textbf v_e}$ associated with the TPD EPWs. For a pair of TPD EPWs,  $\hat{\textbf v}_{e1,2} = \omega_{1,2} {\textbf k }_{1,2} a_{1,2}/(4\pi e n_0 \alpha_{1,2})$ and $\hat{n}_{p1,2} = -k_{1,2}^2 a_{1,2}/(4 \pi e \alpha_{1,2})$ are the enveloped values of electron velocities and electron density modulations, respectively.

In each zone $\textbf{p}_n$ is contributed by two dominant pairs of TPD daughter EPWs which have the same growth rate $\gamma_0$ but are allowed to grow from independent initial conditions. The solution of the three-wave model describing an absolute growth of the two pairs of daughter EPWs in the $n$th zone is $a_{1}^{I,II}=\tilde{a}_{n}^{I,II} \exp(\gamma_0 t)$ and $a_{2}^{I,II}=\tilde{a}_{n}^{*I,II} \exp(\gamma_0 t)$, where the superscripts (I,II) identify which pair, $\tilde{a}_{n}$ are determined by the initial conditions ie. the local thermal noises. As the dominant EPWs at zones on $r_n=r_0$ have the same values of $\gamma_0$, $\omega_{1,2}$, and $k_{1,2}$, the leading term of the angular momentum per unit length in $z$ can be readily obtained by summing up all the zones on $r=r_0$ as

\begin{equation}
	J_z
	= -\frac{l \omega_1 \omega_2}{2\pi\omega_{pe}^2} \frac{ \exp(2\gamma_0 t)}{k_{\bot} v_{0} |\frac{\omega_2}{k_2^2}-\frac{\omega_1}{k_1^2}|} \sum_{n, \\ r=r_0} [\tilde{a}_n^{I}\tilde{a}_n^{*I} + \tilde{a}_n^{II}\tilde{a}_n^{*II} ] \Delta \sigma_n,
	\label{eq:jz}
\end{equation}
where $\omega_{pe}=\sqrt{4\pi n_0 e^2/m_e}$ is the electron plasma frequency and $ \Delta \sigma_n$  is the cross-sectional area in the $x-y$ plane of $S_n$. $J_z$ is found proportional to $l$, showing the dependence of the EPWs' angular momentum on the OAM of the pump laser. Eq.(\ref{eq:jz}) also predicts an exponential growth of $J_z$ with a growth rate $2\gamma_0$, which has been verified by our simulations and plotted in Fig.\ref{fig:compare}(a) by the green squares. $J_z$ of EPWs in the simulations is calculated by integrating ${\textbf r} \times m_en_p{\textbf v_e}$ over the $x-y$ plane of the simulation domain and then averaging it over $z$. 

To bridge the angular momentum carried by the TPD EPWs and by the OAM pump laser, a quantity Angular momentum Over Energy ($AOE$) is introduced for both the EPWs and the pump: $AOE \equiv J_z/W$,
where $W$ is the waves' energy per unit length in $z$. It was found that $AOE_{pump}=-l/\omega_0$ for a linearly-polarized LG laser\cite{Allen2000}, which is consistent with the fact that this LG light has an OAM of $-l\hbar$ per photon. For the TPD EPWs, in our theory the leading term of the EPWs' total energy can be expressed as a sum 
of the dominant EPWs' energies in all of the zones on $r=r_0$, ie. $W_{EPW} = \sum_n (w_{n1}^I+ w_{n2}^I+ w_{n1}^{II}+w_{n2}^{II}) \Delta \sigma_n$, where  $w_{n1,2}^{I,II} = \omega_{1,2}^2k_{1,2}^2  a_{1,2}^{I,II}a_{1,2}^{*I,II} /(8\pi\omega_{pe}^2 \alpha^2_{1,2})$ [reformed Eq.10 of Ref.\cite{Bliokh2022}] is the energy density of the two pairs of dominant EPWs respectively in the $n$th zone. Using Eq.(\ref{eq:jz}) one can readily obtain $AOE_{EPW} = -l/\omega_0$, same as $AOE_{pump}$. In our simulations, the values of $AOE_{EPW}$ averaged over the whole simulation box are plotted in Fig.\ref{fig:compare} (b) and they precisely match the theoretical values.
Identical $AOE$ values for the pump laser and the TPD EPWs (ie. $AOE_{pump}=AOE_{EPW}$) guarantee angular momentum conservation during the TPD process since energy is known to be conserved when passed from the pump to the daughter EPWs given the frequencies' matching condition satisfied. 

The TPD EPWs collectively form a spiral current tube [see Fig. \ref{fig:scope}(a)] that generates collimated axial magnetic fields ($B_z$) inside the tube. Precise modeling on the evolution of $B_z$ requires kinetic simulations. Here we give an estimate on the quasi-static $B_z$ based on the assumption that TPD is saturated such that the time-averaged azimuthal current density $\mathbb{j}_\phi$ reaches a quasi-static value. Then $B_z$ can be calculated as 
$B_z = (4\pi/c)\int_{r}^{\infty}  \mathbb{j}_\phi  dr' $\cite{Shi2018}, where $\mathbb{j}_\phi$ is correlated with $J_z$ as $J_z = (-m_e/e)\int_0^{2\pi}d\phi \int_{0}^{\infty} r^2 \mathbb{j}_\phi  dr$. Under the approximation that $\mathbb{j}_\phi$ is uniformly distributed on a thin tube around $r=r_0$, the maximum ($B_{max}$) of $B_z$ reached inside the tube can be expressed as  $B_{max} = -2 e J_z / (m_e c r_0^2) \approx e l D_{tube} \bar{E}_{EPW}^2/(4m_e c \omega_p r_0)$, where $D_{tube}$ is the thickness of the current tube and $\bar{E}_{EPW}$ is the averaged electric field amplitude due to EPWs on the tube. $D_{tube}$ can be approximated by the full-width-half-maximum of the LG laser amplitude peak at $r=r_0$, ie., $D_{tube}\approx 1.2w_{b0}$. The saturated level of $\bar{E}_{EPW}$ is determined by complicated nonlinear effects absent in our fluid simulations and is still an open question. However, based on our kinetic simulations on plane-wave laser driven TPD \cite{Yan2009,Yan2012,Yanthesis2012}, the saturated $\bar{E}_{EPW}$ due to TPD EPWs were found larger than the laser electric fields in most simulations. So a lower limit of $\bar{E}_{EPW}$ is arguably estimated as $\omega_0 A_{max}/c$, which yields an estimation on $B_{max}$ formulated in practical units.
\begin{equation}
	B_{max} \gtrsim 0.7 I^{max}_{14}\lambda_{\mu m}\sqrt{l}\, [\rm{T}],
	\label{eq:bmax}
\end{equation}
In the case of $l=6$,  Eq. (\ref{eq:bmax}) predicts generation of about 40T quasi-static axial magnetic field via this LG-laser driven TPD process, same order of magnitude as predicted by the theories in Refs.\cite{Shi2018} and \cite{Nuter2020} in the relativistic laser intensity regime. Our mechanism also demonstrates pure laser-driven generation of axial magnetic fields with amplitudes that are similar to the method using a pulsed-power-driven coil\cite{Moody2022}.  




\acknowledgments{This research was supported by the Strategic Priority Research Program of Chinese Academy of Sciences, Grant No. XDA25050400 and XDA25010200, by National Natural Science Foundation of China (NSFC) under Grant Nos. 12175229 and 11621202, by the Fundamental Research Funds for the Central Universities. The numerical calculations in this paper have been done on the supercomputing system in the Supercomputing Center of University of Science and Technology of China.}

\end{document}